\documentstyle[11pt,paspconf,epsf] {article}
\begin{document}

\title{Radiation--Driven Warping: The Origin of Warps and Precession
in Accretion Disks}

\author{Philip R. Maloney} 
\affil{Center for Astrophysics and Space
Astronomy, University of Colorado, Boulder, CO 80309-0389}

\author{Mitchell C. Begelman\altaffilmark{1}}
\affil{JILA, University of Colorado, Boulder, CO 80309-0440}

\altaffiltext{1}{Also at Department of Astrophysical, Planetary, and
Atmospheric Sciences, University of Colorado, Boulder.} 

\begin{abstract}

A geometrically thin, optically thick, warped accretion disk with a
central source of luminosity is subject to non-axisymmetric forces due
to radiation pressure; the resulting torque acts to modify the warp.
Initially planar accretion disks are unstable to warping driven by
radiation torque, as shown in a local analysis by Pringle (1996) and a
global analysis of the stable and unstable modes by Maloney, Begelman,
\& Pringle (1996).  In general, the warp also precesses.

We discuss the nature of this instability, and its possible
implications for accretion disks in X-ray binaries and active galactic
nuclei.  Specifically, we argue that this effect provides a plausible
explanation for the misalignment and precession of the accretion disks
in X-ray binaries such as SS 433 and Her X-1; the same mechanism
explains why the maser disk in NGC 4258 is warped.

\end{abstract}


\section{Introduction}

For more than two decades, observational evidence has 
accumulated for the presence of tilted, precessing accretion disks in
X-ray binary systems. The initial discovery of a 35-day period in the
X-ray flux from Her X-1 (Tananbaum et al. 1972) was interpreted almost
immediately as the result of periodic obscuration by a precessing
accretion disk which is tilted with respect to the plane of the binary
system (Katz 1973). Katz proposed that the precession was forced by
the torque from the companion star, but left unexplained the origin of
the disk's misalignment. (The fact that the optical light from the
companion star HZ Her --- which is largely reprocessed X-ray flux --- 
does not vary significantly over the 35-day period indicates that the
{\it intrinsic} X-ray luminosity is not varying.) An alternative
possibility was proposed by Roberts (1974), who suggested that it was
actually the companion star that was misaligned and precessing; the
material in the accretion disk would be ``slaved'' to the precession
of the companion star provided the residence time in the disk was
sufficiently short.

Interest in tilted, precessing accretion disks was greatly stimulated
by the discovery of the precessing relativistic jets in SS 433 (for a
review see Margon 1984). The systematic velocity variations of the
high-velocity lines, the morphology of the radio jets, and optical
photometry of the system all indicate that the accretion disk is
precessing with a period of 164 days; both the inner region of the
disk, near the compact object (to account for the jet precession) and
the outer edge (to account for the optical photometry) must precess at
the same rate, indicating {\it global} precession of the disk. Global
precession of the disk is also indicated for Her X-1, to explain the
systematic variation of the pulse profiles over the 35-day cycle
(Tr\"umper et al. 1986; Petterson, Rothschild, \& Gruber 1991).

A number of other X-ray binaries, both high and low mass, show
evidence for long period variations and precessing, inclined accretion
disks: LMC X-4 (30.5 days), Cyg X-1 (294 days), XB1820-303 (175 days),
LMC X-3 (198 or 99 days), and Cyg X-2 (77 days) (Priedhorsky \& Holt
1987, and references therein; Cowley et al. 1991; Smale \& Lochner
1992; White, Nagase, \& Parmar 1995, and references therein). Given
the small number of X-ray binaries for which adequate data exist to
test for the existence of such periodicities, it is evident that
precessing, inclined accretion disks are in fact common in X-ray
binaries. Further evidence for the existence of non-planar accretion
disks, in a very different context, is provided by the extraordinary
warped disk of water masers in NGC 4258 (Miyoshi et al. 1995;
Herrnstein, Greenhill, \& Moran 1996).

In spite of twenty years of theoretical work, however, there is no
accepted model for producing and maintaining a steadily precessing,
tilted accretion disk. The initial models for Her X-1 both suffer
from severe flaws. The model of Katz (1973) requires that the outer
disk be tilted with respect to the binary plane, but has no mechanism
for producing such a tilt; furthermore, the strong ($\propto r^{3/2}$)
radial dependence of the quadrupole forcing term implies that the
precession can only be driven at large radius, and the free precession
timescale at the outer edge of the disk is too short compared to the
observed period. This shortcoming was overcome by Merritt \& Petterson
(1980), who included the effects of viscous torques in the disk and
showed that it was possible to obtain global modes with precession
rates slower than the free precession time at the outer edge. However,
the tilt of the disk at the outer edge was still simply imposed as a
boundary condition, with no physical explanation. The slaved disk
model of Roberts (1974) requires that the rotation axis of the
companion star be inclined to the binary plane; however, the axial
tilt will decay by tidal damping faster than the orbit will
circularize (Chevalier 1976; Papaloizou \& Pringle 1982), and the
observed periods are too short compared to the expected periods for
synchronously rotating binary systems (Hut \& van den Heuvel
1981). Furthermore, precession of the companion star is expected to
produce a more precise clock than is seen in Her X-1, for which the
inferred precession period varies by up to ten percent (Boynton, Crosa, 
\& Deeter 1980). Other suggested mechanisms, such as precession of an
oblate neutron star (Brecher 1972) also suffer from the fact that, at
least for Her X-1 and SS 433, a {\it global} precession mode is
required, and it is very difficult to communicate a fixed precession
rate through a viscous, differentially-rotating disk.

\section{Radiation-Driven Warping}

Fortunately, there is a natural mechanism for producing warping and
global precessing modes in centrally-irradiated accretion disks:
namely, warping of disks due to radiation pressure from the central
source.

\begin{figure}
\plotfiddle{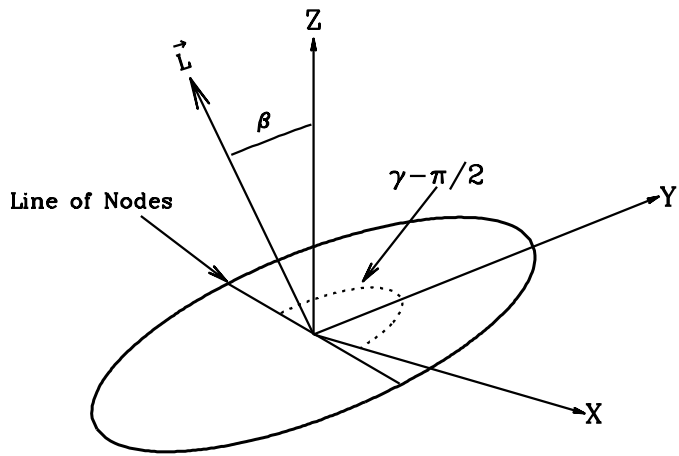}{2.5in}{0}{85}{85}{-295}{-325}
\caption {Geometry of a tilted annulus of an accretion
disk. The equatorial plane of the binary system is the $XY$ plane; the
ring is tipped by an angle $\beta$ relative to the normal to the
equatorial plane (the $Z-$ axis). Also marked is the line of nodes; the
descending node makes an angle of $\gamma-\pi/2$ with the $X-$axis. The
angular momentum vector of the ring is $\vec {\rm L}$.}
\label{fig-1}
\end{figure}

Consider an annulus of an accretion disk with a central source of
luminosity (Figure~\ref{fig-1}). 
If we warp this annulus slightly, it will see the central source of
radiation, and, if the disk is optically thick, it will absorb the
incident flux. The momentum associated with this absorbed incident
flux is purely radial, and so the force due to the {\it incident}
radiation will produce zero torque. If the disk is optically thick to
{\it re-emission} of the absorbed radiation, however, the net
re-emitted flux will be {\it normal} to the local plane of the
disk. The radiation pressure acting on the irradiated surface due to
the re-emitted radiation will therefore exert a torque at each exposed
point of the annulus. If the annulus is warped, i.e., the local angle
of tilt varies with position, then the net torque on the ring will be
non-zero, because the deposition of radiation momentum per unit area
of the ring will be non-axisymmetric. In general, the net torque will
be non-zero provided that there is a radial gradient in either the
tilt angle $\beta$ (in which case the ring will precess) or the angle
of the line of nodes (in which case the torque will cause the tilt to
change); see Figure~\ref{fig-1}. Note that the disk must be
optically thick both to absorption of the incident radiation and to
re-radiation of the absorbed flux. If the disk is optically thin to
{\it absorption}, then the absorption rate will be essentially
constant through the annulus, and the flux re-radiated from the two
faces of the annulus (the irradiated face and the opposite face, which
does not see the central source directly) will be the same, so that
the re-emitted radiation will exert zero torque. If the disk is
optically thin to {\it re-emission}, then the re-emitted radiation
will be isotropic, and again the torque will be zero.

The importance of radiation pressure for altering the shape of {\it a
priori} warped accretion disks was first pointed out by Petterson
(1977). Detailed numerical models of warped disks were constructed by
Iping \& Petterson (1990), while disk models including a related
effect --- the torque from a non-axisymmet\-ric wind driven off the
surface of a disk by X-ray heating --- were presented by Schandl \&
Meyer (1994). Iping \& Petterson, in particular, suggested that
radiation pressure forces might be very important in establishing the
shape of the disk and the precession rate.

This hint that accretion disks might actually be unstable to warping
due to radiation pressure was confirmed by Pringle (1996), who showed
via a local (WKB) analysis that isothermal $\alpha-$disks (radial
surface density $\Sigma\propto R^{-3/2}$), even if initially planar,
are unstable to radiation-pressure warping. However, a
short-wavelength WKB analysis is only marginally justified, as the
unstable modes in Pringle's analysis have wavenumber
$k\sim2\pi/R$. Furthermore, it is impossible to specify the boundary
conditions in a purely local analysis, and it turns out that the
boundary conditions play a crucial role.

Maloney, Begelman, \& Pringle (1996) extended Pringle's work in a
global analysis, and solved the isothermal case analytically. For the
inner boundary condition we require that no viscous torque be exerted
on the disk at the origin; this requires that the local angle of tilt
$\beta\rightarrow\ $constant as $R\rightarrow 0$, so that the disk
becomes asymptotically flat (but not necessarily untilted) at small
radii. The outer boundary condition is that $\beta\rightarrow 0$ at
some finite radius. Physically, this outer boundary condition is
required because the disk material is supplied in the equatorial plane
at the disk's outer edge; therefore, whatever the actual shape of the
warped disk, it must return to the equatorial plane of the system at
the accretion disk boundary.

\begin{figure}
\plotfiddle{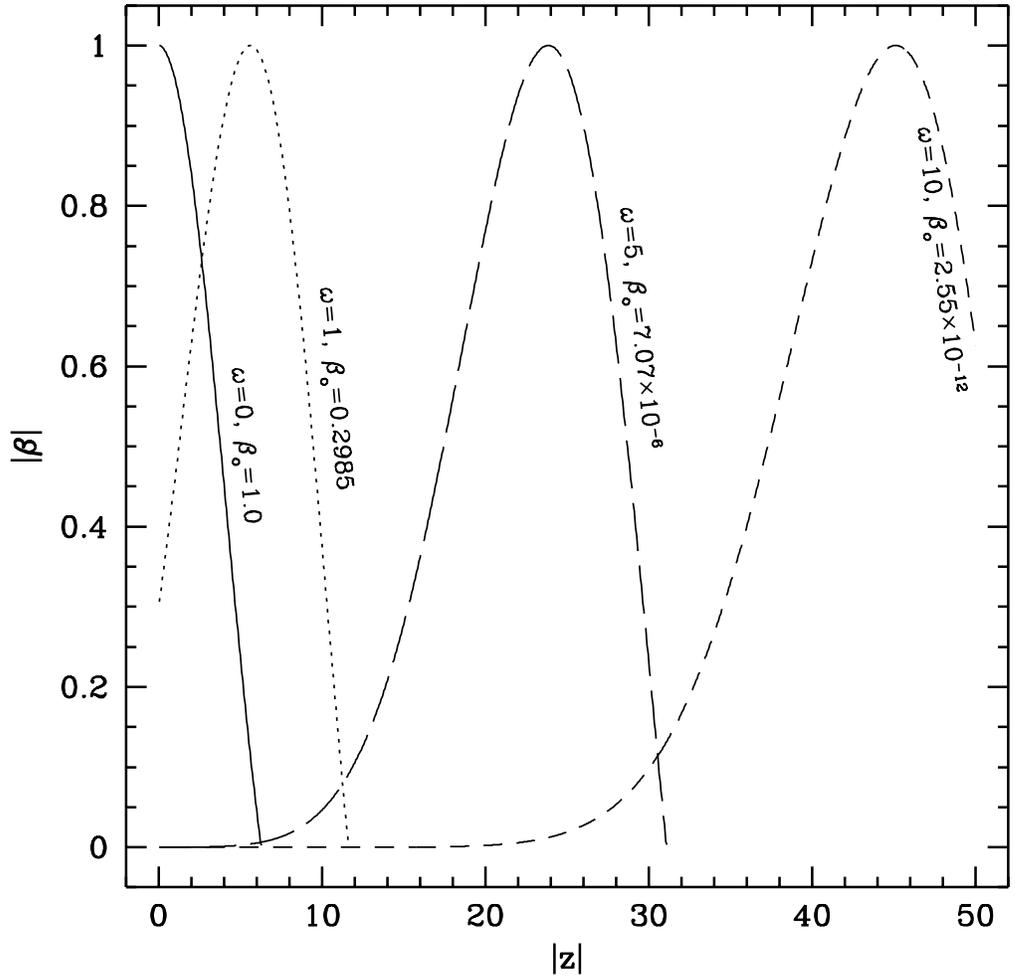}{5.5in}{0}{95}{95}{-287}{-200}
\caption {The amplitude of the local angle of tilt $|\beta|$ for
several of the growing solutions of the twist equation (for an
isothermal disk) is plotted against the radius variable $|z|$, which
is proportional to the radiative efficiency $\epsilon$ times
$(R/R_s)^{1/2}$, where $R_s$ is the Schwarzschild radius. For clarity
the solutions have been plotted only to the location of the first zero
for each mode. All the modes have the same precession rate and grow as
$e^{\omega t}$. They have all been normalized so that the maximum
amplitude is 1; each curve is labeled with the dimensionless growth
rate $\omega$ and with the normalization constant $\beta_o$, i.e., the
magnitude of the tilt at the origin. For $\omega\gg 1$ this is
extremely small, so that for modes with high growth rates the disks
remain flat interior to a critical radius. The maximum value of
$\omega$ for a disk is fixed by the largest unstable radius, which is
either the physical edge of the disk or the radius at which it becomes
optically thin.}
\label{fig-2}
\end{figure}

\begin{figure}
\plotfiddle{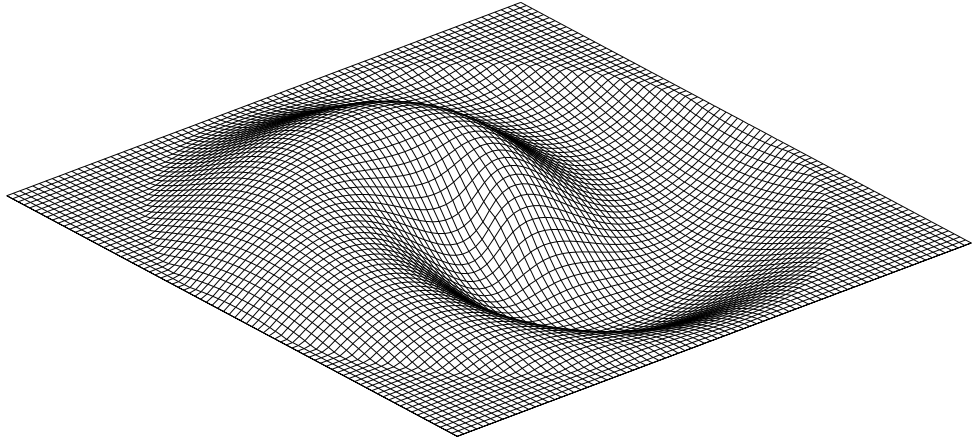}{2.5in}{90}{73}{73}{295}{-110}
\caption {Surface plot of the warp shape for the
steady-state mode of an isothermal ($\delta=3/2$) accretion disk. The
warp has been plotted out to the radius at which it returns to the
equatorial plane; the amplitude of the warp has been fixed at 10\% of
this radius.}
\label{fig-3}
\end{figure}

There are stable, unstable, and time-independent modes for warping of
the disk. The precession rate of the warp is in general
non-zero. There is only a single value of the precession rate (i.e.,
the real part of the frequency eigenvalue) for which the warp of the
disk returns to the equatorial plane {\it at any radius}; this is the
only solution that can be matched to a disk that is unwarped at its
outer boundary. The growth rate of the warp (the imaginary part of the
eigenvalue) can take on a range of values, depending on the outer
boundary condition. The location of the warp and the first zero move
outward with increasing growth rate (Figure~\ref{fig-2}),
so that the warp grows from the ``outside'' in. The warp 
precesses as a fixed pattern by twisting the disk in such a way that
the radiation torque causes each radial annulus to precess at the same
rate. Figure~\ref{fig-3}  
shows a surface plot of the steady-state mode of the
isothermal disk, plotted out to the radius at which the warp
returns to the equatorial plane. The amplitude of the warp has been
fixed at 10\% of the outer radius.

The isothermal case is actually degenerate (the real parts of all the
eigenvalues are identical), but the degeneracy is lifted for
nonisothermal surface density distributions, and a complex pattern of
modes results (Maloney, Begelman, \& Nowak 1996). However, there are a
number of features of the instability which are generic:
\begin{itemize}
\item The radiation torque per unit area is $\Gamma\sim
(L/4\pi R^2c)\times R$, while the angular momentum density is $l\sim
\Omega R^2\Sigma\sim \Omega R^2\dot M/\nu$, where $\Omega$ is the
angular velocity at radius $R$ in the disk, $\dot M$ is the mass
accretion rate, and $\nu$ is the kinematic viscosity. The radiation
torque timescale is thus given by
\begin{equation}
t_{\rm rad}\sim {\Omega R^3\over c\nu}{\dot M c^2\over L} ={\Omega
R^3\over c\nu}\;\epsilon^{-1}
\end{equation}
which, for an accretion-fueled source, depends only on the accretion
efficiency $\epsilon=L/\dot M c^2$ and not on the luminosity $L$ and
the mass accretion rate individually.
\item The viscous timescale is $t_{\rm visc}\sim R^2/\nu$, so
the ratio of viscous to radiation torque timescales is given by
\begin{equation}
{t_{\rm visc}\over t_{\rm rad}}\sim {\epsilon c\over \Omega R}\sim
\epsilon \left({R\over R_g}\right )^{1/2}\;;\quad\quad R_g={GM\over
c^2}
\end{equation}
which depends only on $\epsilon$ and the radius normalized to the
gravitational radius. Thus the radiation torque always wins out at
large radius, i.e., for $R > \epsilon^{-2} R_g$, but viscosity always
dominates near the center. This is why the disk will be {\it flat}
(but in general will have non-zero tilt) at small radius.
\item For a fixed outer boundary radius, there is a single
value of the precession rate; the location of the warp always moves
outward in radius with increasing growth rate. This is why the outer
boundary condition is crucial: the outer boundary radius determines
the growth rate of the warp. This self-regulation means that the
growth time will be of approximately the same order as the precession
time. For X-ray binaries, this timescale is much shorter than the
lifetime of the accretion disks, and so the warps in X-ray binary
systems presumably reach the steady-state solutions. Although the inner
region of the disk ($R \ll \epsilon^{-2} R_g$) will be flat, it will
generally be {\it tilted} with respect to the binary orbit plane, and
{\it the entire disk shape} will precess globally, at the same
rate. Because it is set by conditions in the disk, the precession
timescale may not be precisely constant with time, but will be coupled to
variations in the outer boundary radius, the disk surface density
distribution, and the accretion efficiency.
\end{itemize}

\section{Applications}

\subsection{X-ray Binaries}

The precession timescale due to radiation pressure can be written as
\begin{equation}
t_{\rm prec}\sim 32 \left({\alpha\over
0.1}\right)^{-1}\left({\epsilon \over 
0.1}\right)^{-1} \left({h/R\over 0.01}\right)^{-2} R_{11}\;{\rm days}
\label{precession} \end{equation}
where $\alpha$ is the usual disk viscosity parameter, $\epsilon=L/\dot
M c^2$ is the radiative efficiency (assuming an accretion-fueled
source), $h/R$ is the thickness/radius ratio of the disk at the
location of the warp, and the outer radius of the disk is at
$R=10^{11}R_{11}$ cm; we have normalized to typical values for an
X-ray binary. This is clearly of the right order to explain the
observed precession periods.

In X-ray binaries, however, the quadrupole torque due to the companion
star can also have a significant influence on the warp and its
precession rate (e.g., Merritt \& Petterson 1980). In the absence of
such a torque, the precession of the warp can be either prograde or
retrograde with respect to the direction of rotation, due to the
antisymmetry of the net torque direction for inversion of the warp
through the origin. The torque due to the other component of the
binary breaks this symmetry, however, and establishes a preferred
direction for precession, counter to the orbital direction. The
radiation-driven warping mechanism is still essential in this case
because it establishes the warp on which the quadrupole torque
operates. Because of the strong radial dependence of the quadrupole
torque term, there will be a characteristic radius at which the
radiation torque and quadrupole torque will be equal, with the
quadrupole term rapidly dominating at larger $R$ and the radiation
term at smaller $R$. Because of this dominance at large $R$, if the
quadrupole term is important it will always set the boundary condition
at the outer edge of the disk, thereby controlling the resulting
precession rate.

\subsection{The Warped Maser Disk of NGC 4258}

VLBI observations of the 22 GHz water maser line emission from the
nucleus of NGC 4258 (which possesses a low-luminosity active nucleus)
have shown that the maser emission arises in a thin, warped disk, with
an inner radius of 0.13 pc and an outer radius of 0.25 pc, which
exhibits a perfectly Keplerian velocity curve to within the
measurement errors (Miyoshi et al. 1995; Greenhill et al. 1995). The
derived central mass is $M_c=3.6\times 10^7 M_\odot$, and dynamical
arguments essentially rule out mass models other than a supermassive
black hole (Maoz 1995). The obliquity of the warp is $\mu\approx 0.25$
at the outer edge of the masing zone (L. Greenhill 1995, private
communication) and decreases steadily inward.

Neufeld, Maloney, \& Conger (1994) showed that X-ray irradiation of
dense molecular gas is an effective mechanism for generating powerful
water maser emission. By modeling the NGC 4258 disk using the standard
$\alpha-$prescription for viscosity, Neufeld \& Maloney (1995) showed
that the disk would be extremely thin ($h/R\ll 10^{-2}$), that the
transition from molecular to atomic gas should be identified with the
outer edge of the masing region, and that the mass accretion rate
through the disk is $\dot M/\alpha\approx 7\times 10^{-5} \; M_\odot
\;{\rm yr}^{-1}$. (See also Maloney, this volume.) This mass accretion
rate in turn implies that the 
radiative efficiency $\epsilon\approx 0.1$. The existence of the warp
in this picture is crucial, since for a flat disk neither the greatly
reduced irradiation from the central source nor heating by viscous
dissipation would keep the disk warm enough to produce substantial
amounts of water or excite the masing transitions. In fact, Neufeld \&
Maloney (1995) suggested that the inner edge of the masing disk marked
the inner boundary of the warp, as is the case in the model for the
warp derived by Miyoshi et al. (1995). Although an arbitrary imposed
warp would be kinematically stable in a Keplerian potential, since the
vertical and azimuthal frequencies are the same, the origin of the
warp in NGC 4258 remained puzzling.

The radiation-driven warping instability provides a plausible
explanation for the warping of the disk (Maloney, Begelman, \& Pringle
1996). The isothermal models depicted in Figure 2 are applicable to
this case and, given the estimated locattion of the warp and the
accretion parameters derived by Neufeld \& Maloney (1995), we estimate
the dimensionless growth rate to be $\omega \approx 10$.  It is likely
that $\omega_{\rm min}$ is not much smaller than this value, due to
the limit set by the finite age of the accretion disk. The
characteristic timescale for growth of the instability is
\begin{equation}
\tau_i(\omega)\sim{4\over \omega}{c R_s\over \alpha c_s^2\epsilon}
\approx 3\times 10^7 {M_8\over \omega \alpha T_3 \epsilon_{0.1}}\;{\rm
yrs} 
\end{equation}
where $M_c=10^8 M_8 \ M_\odot$, $c_s$ is the sound speed, the gas
temperature $T=10^3 T_3$ K, and the numerical coefficient is for
molecular gas. In the masing molecular zone, $0.3 < T_3 < 1$. While
the characteristic growth time for the $\omega=10$ mode in NGC 4258 is
$\tau_i\sim 5\times 10^6$ years, the time required for even the
$\omega=5$ mode to become comparable in amplitude is an order of
magnitude longer. Thus we expect that the warp will be dominated by
modes with $\omega$ of order 10, unless the accretion
disk is extremely long-lived ($\tau_{\rm disk}\gg 10^8$ years). In
contrast to the X-ray binaries, the warp in NGC 4258 is far from
attaining a steady state.   

Note that the predictions of the radiation-driven warp instability
model are in reasonable agreement with the observations of NGC 4258,
provided that $\epsilon\sim 0.1$.  This instability cannot provide an
explanation for the warp if NGC 4258 possesses an advection-dominated
accretion disk with $\epsilon \sim 10^{-3}$, as suggested by Lasota et
al. (1996). This would require $\omega < 1$, implying a prohibitively
long growth timescale, with $\tau_i > 10^9$ years.

\section{Conclusions}

Warping and precession driven by radiation pressure offers a robust
mechanism for producing tilted, precessing accretion disks in
accreting binary systems such as Her X-1 and SS 433, and warped disks
in AGNs. Because the warp adjusts itself in such a way as to precess
at a uniform rate at all radii, with the rate fixed by the outer
boundary condition, radiation-driven warping is an inherently global
mechanism, thereby avoiding the difficulties inherent in other
proposed mechanisms for producing warping and precession. This
mechanism can thus explain the simultaneous precession of inner disks
(as evidenced by the jets of SS 433 and the pulse profile variations
of Her X-1) and outer disks (as required to match the periodicities in
X-ray flux and disk emission in these same objects).

Our work to date has been based on a linearized version of the
equations governing the evolution of a twisted, viscous accretion disk
(e.g., Pringle 1992). Although this is adequate for establishing the
nature of the instability, it leaves unanswered the question of the
end state of the instability, i.e., at what amplitude the warp will
saturate. In addition, there are a number of other complications which
we have so far ignored, including torquing by a companion, torques due
to X-ray heated winds (as considered by Schandl \& Meyer 1994), and
opacity effects due to material intervening between the central
radiation source and the disk, particularly the shadowing of the outer
parts of the disk by the warp itself. It is also important to
establish the observational consequences of radiation-driven warping;
these will include not only the shape and precession rate of the disk,
but the effect of the self-irradiation on the thermal state of the
disk.  Answering any of these questions will almost certainly require
three-dimensional numerical simulations, but these are now well within
reach of modern computing capabilities.

\acknowledgments 

We thank Jim Pringle --- the discover of the radiation-driven warp
instability --- for his constant encouragement and his active
collaboration in some of the work described above.  We also
acknowledge support from NSF grants AST-9120599 and AST-9529175, and
from NASA Long-Term Astrophysics Program grant NAGW-4454. Finally, we
note with sorrow the recent passing of Jacobus Petterson, who
pioneered the study of warped accretion disks.

\end{document}